\pdfoutput=1
\documentclass[]{spie}  

\usepackage{url}
\usepackage{amsmath,amsfonts,amssymb}
\usepackage{graphicx}
\usepackage[colorlinks=true, allcolors=blue]{hyperref}
\usepackage{enumitem}
\setlist[itemize]{noitemsep, nolistsep}
\usepackage{hyperref}
\usepackage{aas_macros}
\usepackage{authblk}

\usepackage[british]{babel}

\title{GRINTA: a new Probe of the Dynamic Universe}

\begin{document}

\author[a]{Lorenzo Natalucci}
\author[b]{Philippe Laurent}
\author[c,d]{Marica Branchesi}
\author[a]{Mariateresa Fiocchi}
\author[e]{Lorraine Hanlon}
\author[f]{J. Miguel Mas-Hesse}
\author[b]{Aline Meuris}
\author[g]{Paul ‘O Brien}
\author[a]{James C. Rodi}
\author[h]{Simone Scaringi}
\author[i]{Konrad Skup}
\author[j]{Norbert Werner}
\author[k]{Silvia Zane}
\author[f]{Julia Alfonso-Garzon}
\author[c]{Biswajit Banerjee}
\author[a]{Patrizia Barria}
\author[l]{Anthony J.Bird}
\author[m]{Soren Brandt}
\author[a]{Gabriele Bruni}
\author[n]{Floriane Cangemi}
\author[o]{Gianluca Castignani}
\author[n]{Eric Chassande-Mottin}
\author[n]{Sylvain Chaty}
\author[m]{Jerome Chenevez}
\author[n]{Alexis Coleiro}
\author[p]{Francesco Dazzi}
\author[c]{Alessio L. De Santis}
\author[g]{Phil Evans}
\author[q]{Olivier Godet}
\author[b]{Diego Gotz}
\author[r] {Rene Hudec}
\author[s]{Lucien Kuiper}
\author[o]{Angela Malizia}
\author[e]{Sheila McBreen}
\author[t]{Sandro Mereghetti}
\author[t]{Manuela Molina}
\author[j]{Filip Münz}
\author[e]{David Murphy}
\author[c,d]{Gor Oganesyan}
\author[a]{Francesca Panessa}
\author[o]{Elena Pian}
\author[j]{Jakub \v{R}\'{\i}pa}
\author[u]{Oliver J. Roberts}
\author[c]{Samuele Ronchini}
\author[b]{Fabian Schussler}
\author[r]{Vojtěch Šimon}
\author[g]{Nial Tanvir}
\author[e]{Alexey Uliyanov}
\author[a]{Nello Vertolli}
\author[r]{Stanislav Vìtek}
\author[a]{Ugo Zannoni}

\affil[a]{INAF/IAPS, Via del Fosso del Cavaliere 100, 00133 Rome, Italy }
\affil[b]{CEA Commissariat à l'Energie Atomique Saclay, 
Gif-Sur-Yvette Cedex, France }
\affil[c]{ Gran Sasso Science Institute (GSSI), I-67100 L’Aquila, Italy }
\affil[d]{ INFN, Laboratori Nazionali del Gran Sasso, I-67100 Assergi, Italy }
\affil[e]{ Centre for Space Research, University College Dublin, Dublin 4, Ireland}
\affil[f]{ Centro de Astrobiología (CAB), CSIC–INTA, Villanueva
de la Cañada, Spain}
\affil[g]{School of Physics \& Astronomy, Univ. of Leicester, University Road, Leicester, LE1 7RH, UK }
\affil[h]{ Centre for Extragalactic Astronomy, Department of Physics, Durham University, South Road, Durham DH1 3LE, UK}
\affil[i]{Centrum Badan Kosmicznych, Polish Academy of Science, Bartycka 18a, 00-716 Warszawa, Poland }
\affil[j]{ Department of Theoretical Physics and Astrophysics, Faculty of Science, Masaryk University, Kotl\'a\v{r}sk\'a 2, Brno 611 37, Czech Republic }
\affil[k]{ Mullard Space Science Laboratory, University College London, Holmbury St Mary, Dorking, Surrey RH5 6NT, UK }
\affil[l]{School of Physics and Astronomy, University of Southampton, Southampton SO17 1BJ, UK }
\affil[m]{DTU-Space, Technical University of Denmark, Lyngby, Denmark}
\affil[n]{ Universit´ e Paris Cit´ e, CNRS, CEA, Astroparticule et Cosmologie, F-75013 Paris, France}
\affil[o]{INAF-Osservatorio di Astrofisica e Scienza dello Spazio di Bologna, via Gobetti 93/3, I-40129, Bologna, Italy}
\affil[p]{INAF, viale del Parco Mellini 84, I-00136 Rome, Italy }
\affil[q]{IRAP, Universite de Toulouse, CNRS, CNES, Toulouse, France }
\affil[r]{Faculty of Electrical Engineering, Czech Technical University,16636 Prague, Czech Republic }
\affil[s]{SRON Space Research Organization Netherlands, Niels Bohrweg 4, 2333 CA Leiden, The Netherlands }
\affil[t]{INAF–Istituto di Astrofisica Spaziale e Fisica Cosmica di Milano, via A. Corti 12, I-20133 Milano, Italy}
\affil[u]{Physics, School of Natural Sciences, University Road, University of Galway, Galway, H91 TK33, Ireland}

\authorinfo{Further author information: (Send correspondence to L. Natalucci)\\L. Natalucci: E-mail: lorenzo.natalucci@inaf.it, Telephone: +39 06 45488 461 \\  
}


\maketitle

\newcommand\spiecopyright{%
  \begingroup
  \renewcommand\thefootnote{}\footnote{Copyright 2026 Society of Photo-Optical Instrumentation Engineers (SPIE). One print or electronic copy may be made for personal use only. Systematic reproduction and distribution, duplication of any material in this publication for a fee or for commercial purposes, and modification of the contents of the publication are prohibited. DOI: \url{https://doi.org/10.1117/12.3105908}}%
  \addtocounter{footnote}{-1}%
  \endgroup
}
\spiecopyright

\begin{abstract}

Recent observations of the transient sky at all wavelengths are increasingly revealing the importance of the multi-messenger and multi-wavelength approach. The \textit{GRINTA} (Gamma-Ray INternational Transient Array Observatory) mission, proposed for launch around the middle of the next decade, is conceived as a small mission with good sensitivity, excellent angular resolution and fast follow-up capability for studying transient sources at timescales from ms to hours, at the same time ensuring optimal integration with multi-messenger networks. The \textit{GRINTA} mission will carry two complementary payloads to cover in total the 5~keV-10~MeV band, that will detect and localise gamma-ray bursts covering $\sim$~half of the sky and will be able to perform imaging surveys with sub-arcmin resolution. \textit{GRINTA} will operate in synergy with the most powerful electromagnetic, gravitational wave and neutrino observatories foreseen to be operational after $\sim2035$.
\end{abstract}

\keywords{Time-Domain Astronomy, Transient Astrophysics, Multi-Wavelength Follow-up, X-ray Surveys}

\section{INTRODUCTION}
\label{sec:intro}  
The hard X-ray regime is critical to the understanding physics at extreme gravity and temperatures. It also provides a band that is mostly absorption-free and so giving access to the study of cosmic explosions within a large volume of the Universe. These explosions are expected to occur in extreme environments that often produce hard X-rays. 

Electromagnetic (EM) counterparts to multi-messenger (MM) sources like e.g. the ones associated to gravitational waves (GW) and high energy neutrinos provide information key to unveil their underlying physics. Hard X-rays are of outmost importance to reveal and study the prompt emission of these events, mostly associated to the presence of a gamma-ray burst (GRB). 
However, the absence of planned future wide field hard X-ray/gamma-ray missions during the 2030 to 2040 decade is a major challenge to future multimessenger discoveries. At that time, despite the remarkable success achieved by the current fleet of gamma-ray missions, many instruments will have far exceeded their nominal lifespan. 
The impossibility of monitoring prompt and afterglow emissions due to the lack of high-energy observatories in the 2030's will significantly limit our ability to maximise the scientific return of multimessenger sources by detecting an electromagnetic (EM) counterpart. The Gamma-Ray INTernational Array observatory (\textit{GRINTA}) has been proposed to fill this gap. 

\textit{GRINTA} is meant to explore the transient hard-X/gamma-ray sky implementing a \( \sim 8\) steradian FoV soft gamma-ray detector to hunt for transient events in the sky, such as GRBs associated to binary mergers and core-collapse of massive stars, magnetar flares plus other, less known high-energy transients, and EM counterparts of cosmic high energy neutrinos. This instrument will be paired to a highly sensitive hard X-ray imager, for which the main task is to perform a fast follow-up with accurate source locations. The \textit{GRINTA} mission is designed to fly in Low Earth Orbit (LEO) and will have autonomous, fast repointing capability to follow up the impulsive and highly variable emission associated with these events. The Transient Event Detector (TED) has an unprecedented coverage compared with previous all-sky monitors, with a factor \( \sim 2\)  sensitivity better than Fermi/GBM and the Gamma-Ray Monitor on board the Space Variable Objects Monitor (SVOM), launched in 2024. Whilst the coded aperture Hard X-ray Imager (HXI), with its FoV covering \( \approx 400\) deg\(^2\) at $>50$\% sensitivity, 
is well suited to explore the error regions of the GRBs localised by the TED but also by other GRB hunting detectors, that will be possibly launched in the same decade, and last but not least, to follow up alerts from GW signals provided by the ground-based interferometers. 


The \textit{GRINTA}/HXI being mainly conceived for follow-up, is also a formidable tool for sensitive hard X-ray surveys, giving a source location capability of $\sim40^{\prime\prime}$ (a factor $\approx5$ better than its predecessor, the IBIS instrument on board \textit{INTEGRAL}). This will allow to complement the future multi-wavelength sky surveys performed by the most powerful telescopes operating at the same epoch. The estimated sensitivity of HXI is $\approx1$ mCrab in the \(5 -30\) keV range [$\approx6$ mCrab in the \(30 -60\) keV range] for a $10^{4}$~s observation. These surveys will allow  discovery of new sources, and drive the identification of hundreds of unidentified gamma-ray sources, which is one of the important heritages of the \textit{Swift}, \textit{INTEGRAL}, and \textit{Fermi} missions. These observations will provide insight into the nature of a wide array of objects, extra-galactic and galactic. 

A summary description of \textit{GRINTA} is given in Table\ref{tab:summary}.
\begin{table}[!h]
        \centering
        \caption{Summary of \textit{GRINTA} characteristics}
        \label{tab:summary}
        \scalebox{0.73}{
        \begin{tabular}{| c| p{19.5cm}|}
        \hline
        Key Science Goals & \begin{itemize}[noitemsep]
            \vspace{-2mm}
            \item Understand the physics of mergers responsible for emission of gravitational waves 
            \item Probe the nature of jets and structure in gamma-ray bursts
            \item Understand the physical processes driving the high energy transient phenomena and clarify their relationship with multi-messenger sources
            \item Understand the physics of compact objects and characterise their populations (surveys)
        \end{itemize}     \\ 
        \hline
          Payloads           & Two instruments:
        \begin{itemize}[noitemsep]
            \item Transient Event Detector (TED), \(0.02-10\) MeV, FOV \(\sim 8\) ster
            \item Hard X-ray Imager (HXI), \(5-200\) keV, FoV \( \approx 400\) deg\(^2\) at $>50$\% sensitivity, location accuracy \( \sim 40\)’’ (SNR=8)
        \end{itemize}       \\ \hline
          Mission Profile    &   \vspace{-2mm}
        \begin{itemize}[noitemsep]
            \item Vega-C, Low Earth orbit, \(\sim\) 5deg inclination
            \item Duration: 2-year (nominal) \(+\) 3 years (extended)
            \item Communication links: equatorial GS, intersatellite relay link (e.g. Iridium~NEXT, Inmarsat)
            
        \end{itemize}           \\  \hline
        Spacecraft           &    
        \vspace{-2mm}
        \begin{itemize}[noitemsep]
            \item 3-axis stabilised
            \item Rapid repointing, slew time : 50\(^{\circ}\)/min
            \item Pointing Requirements: APE=1.0', RPE=0.3', PDE=0.3', AKE=0.1'
            \item Power : 490 W
            \item Dry Mass : 307 kg
            \item S-band/X-band (Kourou, optional Malindi) 
        \end{itemize}    \\  \hline
        \end{tabular}
        }
\end{table}
\vspace{2mm}

\section{SCIENTIFIC OBJECTIVES}

\textit{GRINTA}’s large FoV and high sensitivity at hard X-rays make it ideal to fully utilise MM observations and unlock the understanding of these events. This will allow us to clarify the link between the different cosmic messengers (GW, neutrinos, cosmic rays) with the EM emission spanning the full spectrum of frequencies. Among them, events associated to Short and Long GRBs, Fast Radio Bursts (FRBs), Tidal Disruption Events (TDEs), etc. 
\textit{GRINTA} is perfectly suited for these studies in synergy with the major ground based facilities, such as the LIGO/Virgo upgraded network of observatories, KM3NeT and the big radio and optical facilities expected to be operative in the 2030s (e.g. SKAO, ngVLA, Vera Rubin, ELT) as well as high energy facilities (e.g., \textit{newATHENA}, CTAO). 
The \textit{GRINTA} observatory is also a powerful tool to complement the future multiwavelenght surveys performed by the most powerful telescopes operating in the 2030's. The high sensitivity, high angular resolution and wide field-of-view (FOV) of the hard X-ray imager will allow diverse studies. For extra-galactic sources, mapping the emission components of those nearby, exploring the physics of emission through multi-wavelength campaigns, and population study of AGN within a redshift of \(\sim 0.2\), as well as shedding light on the nature of the progenitors for high-energy cosmic rays and FRBs. For Galactic sources, the imager’s same characteristics will allow to investigate the physical mechanisms at work in accretion and ejections in X-ray binaries, magnetars, X-ray bursters, super-giant fast X-ray transients (SFXTs) and TDEs. The all-sky survey will also be used for discovering new sources and find correlations with unassociated sources from catalogues such as \textit{eRosita} at softer X-rays and \textit{Fermi}/LAT at gamma-rays. Also, TED’s high time resolution and large area will allow to study the nature of emission in soft gamma-ray pulsars up to MeV energies. 

\subsection{Gamma-Ray Bursts and Gravitational Wave Physics}
\subsubsection{GRB physics} 
Gamma-ray bursts are one of the most intriguing transients of the Universe whose physical nature continues to pose fundamental questions, and
were recently detected coincident with a gravitational waves source. 
Their relativistic outflows are capable of emitting up to 10$^{53}$ erg in a timescale of seconds in the hard X- and gamma-rays during the 
prompt phase. This phase is followed by an afterglow emission which lasts for hours, days, and months covering the electromagnetic spectrum from 
radio waves to very-high-energy gamma-rays. The GRB luminosity, combined with the Universe transparency to Mev gamma-rays, allows these cosmic lighthouses to be detected out to high redshift, with the record currently being $z=8.2$ (spectroscopic) 
\cite{Tanvir2009}. 
The distribution of the prompt emission duration and the spectral properties shows a bi-modality which was traditionally associated with two classes of GRBs associated to different progenitors (e.g. Ref.\citenum{1993ApJ...413L.101K}); the ‘long’ GRBs lasting $>$2 seconds and the 'short' GRBs lasting $<2$ seconds.  The presence of supernova emission in the region of some ‘long’ GRBs clearly identifies core collapse of massive stars as their progenitor, whereas 
the origin of 'short' GRBs has been associated with the merging of binary NS systems, as proven by the coincident detection of
GRB~170817A and the gravitational wave source GW170817 
\cite{GW170817BNS, GWGRB2017, Goldstein2017, Savchenko2017}. 

Besides the above mentioned classification
other important classes, recently emerging are Ultra-long GRBs (ULGRBs)(Ref.~\citenum{Virgili2013,Levan2014} and refs therein) and long GRBs associated with emission from r-process nucleosynthesis\cite{Rastinejad2022, Levan2023}, so matching the characteristic BNS merger profile. Magnetar giant flares (MGFs) have also been found to produce relativistic jets resembling the short GRB characteristics. 
In the last 50 years, three magnetar giant flares (MGFs) have been observed in the Milky Way and Large Magellanic Cloud~\cite{Mazets1979,Hurley1999,Palmer2005}. Additionally, the initial spikes of several other MGFs have been been detected from extragalactic candidates located in nearby star-forming galaxies out to distances of $\sim$10~Mpc\cite{Svinkin2021,Trigg2024,Mereghetti2024}. These extragalactic MGFs masquerade as about 2~$\%$ of the short GRB population~\cite{Burns2021}, the tail undetectable with monitoring instruments past 3.5~Mpc (M82), except for \textit{NICER} or the \textit{Swift}/XRT~\cite{Trigg2025}. 
  
\begin{figure}[ht]
    \centering
    \includegraphics[width=0.6\linewidth]{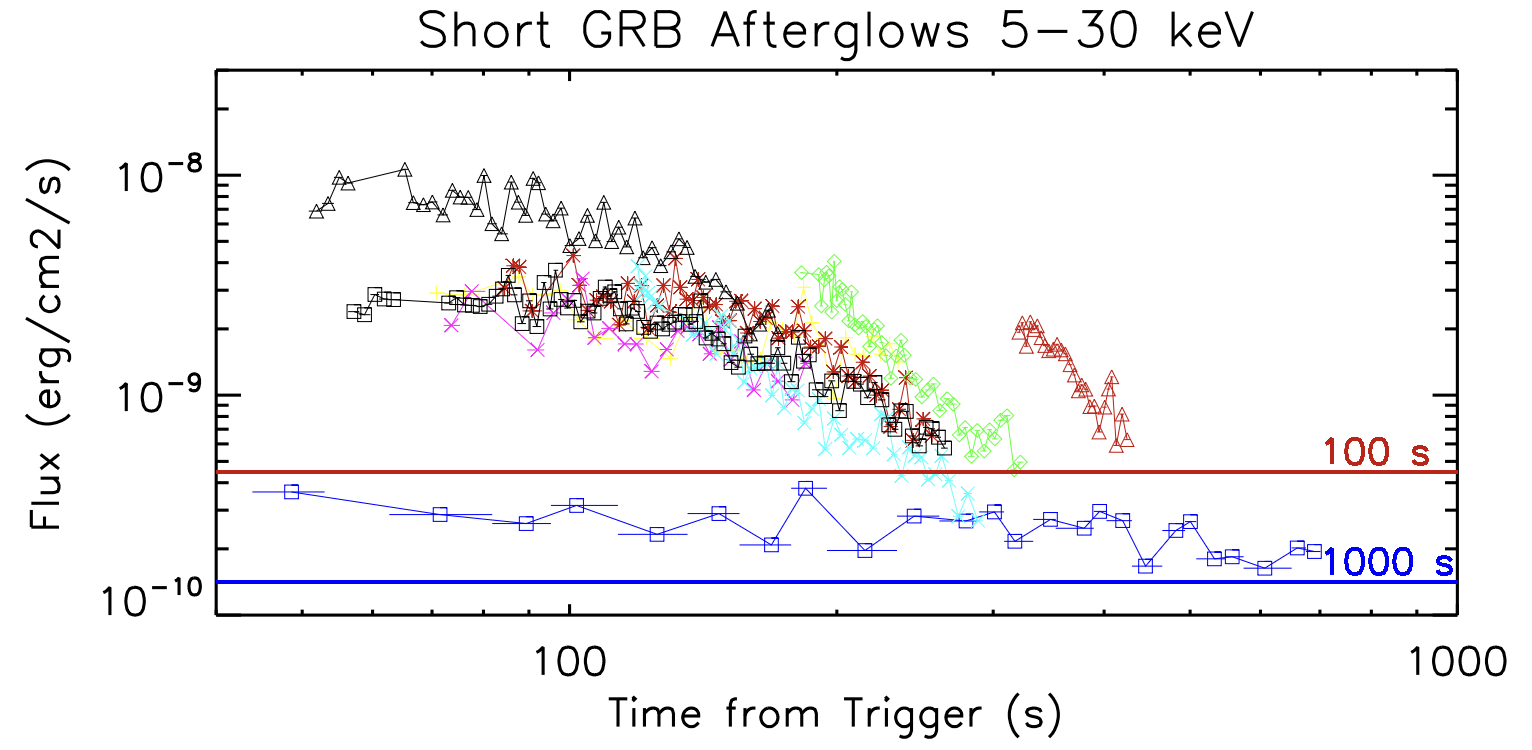}
    \caption{Sample of XRT sGRB afterglows extrapolated to 5--30\,keV. The continuous, red and blue lines are the minimum detectable flux for an exposure of 100s and 1000s, respectively.} 
    \label{fig:afterglow_xrt}
\end{figure}

TED will detect the prompt emission of about 65 short GRBs (sGRBs) and 320 long GRBs (lGRBs) per year, respectively. 
To investigate how many sGRB afterglows HXI would expect to detect, we used the \textit{Swift}/XRT online GRB catalogue \cite{Evans2009}, which currently spans roughly 20 years.  For each sGRB detected by \textit{Swift}/BAT that had an afterglow detected by XRT we extrapolated the best-fit spectral model of the afterglow up to 30 keV to determine the 5--30 keV flux, which we then compared to the predicted HXI sensitivity.  We found that for the sample of 65 \textit{Swift} sGRBs observed on-axis by XRT within 300\,s, the afterglow of \(\sim\)50\% would have been detected above 5\(\sigma\) by HXI.  The light curves for a subset of the events are shown in Fig.~\ref{fig:afterglow_xrt}.  

The source location accuracy of HXI ($\sim 40^{\prime\prime}$) will enable multi-wavelength follow-up by ground-based detectors and the identification of the host galaxies.
Critically, TED and HXI will cover the transition from prompt emission to afterglow in hard X-rays.  This energy range is poorly studied in afterglows and lacks clear spectral characterisation, which is needed to understand the afterglow cooling frequency, which is determined by the characteristics of the circumburst medium.  The position of the cooling frequency depends on whether the medium is made by the collapsar winds or is rarefied in the case of binary neutron star mergers.  Also, the MeV-band is intriguing as MeV anomalies that include double-component structures, possible MeV absorption features \cite{Oganesyan2026}, and early rises of the MeV afterglow \cite{Mohnani2026} have been repeatedly observed.  


\subsubsection{Gravitational-wave counterparts}
In the multi-messenger context, the independent discovery of a gamma-ray burst, GRB~170817A, associated with GW170817 led to
the most extensive follow-up observing campaign in human history\cite{followup_Abbott_2017}, which provided detection of kilonova emission, identification of the host galaxy, and multi-wavelength observations of the GRB afterglow emission \cite{followup_Abbott_2017}. The observed time delay between GRB 170817A and GW170817 provided new insight into
fundamental physics by enabling to constrain the speed of gravity, place new bounds on the violation
of Lorentz invariance, and test the equivalence principle [5, 16]. This event highlighted the importance 
of joint gravitational wave/gamma-ray observations especially in the coming years, when GW observatories will explore more distant volumes of the Universe, where the intrinsically fainter kilonova emission will be hardly detected (at z $> 0.3$) and the only detectable counterparts will come from the high-energy emission. 


The study of a "golden sample" of GW signals is a high scientific priority of \textit{GRINTA}. This can potentially impact on several open questions about progenitor properties, jet physics, and study of the emission mechanisms in relation to the merger remnant; along with gathering information on the properties of the host galaxies and their link to the compact object formation channels.  Depending on the final number of joint events detected, it will be also possible to contribute to cosmology with the Hubble constant measurements using the standard siren method. During the early 30's, the LIGO–Virgo–KAGRA network is expected to include the so-called `post-O5' upgrades, in particular the LIGO A$^{\sharp}$ configuration\cite{Gupta2024}. The next generation of GW observatories, such as the Einstein Telescope (ET) and Cosmic Explorer (CE), will further significantly increase the detection up to $\sim 10^5$ BBH mergers and  $\sim 10^5$ BNS mergers per year \cite{Maggiore2020}. The start of operation of ET and CE is expected sometimes between 2035 and 2040 \cite{GWIC}.


\begin{figure}
    \centering
    \vspace{-2mm}
    \includegraphics[width=0.99\textwidth]{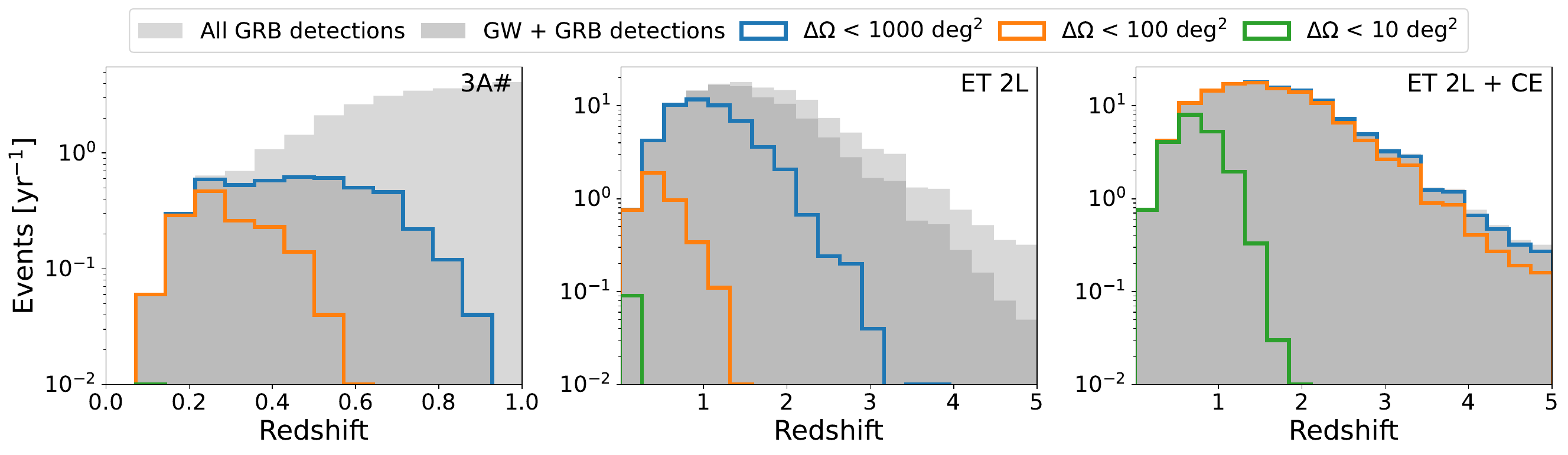}
    \vspace{-2mm}
    \caption{Redshift distribution of the joint GW+$\gamma$-ray detections with \textit{GRINTA}/TED with 3A$^{\#}$ (left), and ET in 2L configuration (center) and ET 2L + Cosmic Explorer (right). The histograms are normalised to one year of observations. $\Delta \Omega$ indicates the area of the GW sky localisation.}
    \label{z_hist}
\end{figure}

In order to evaluate the joint detection rates for GW and gamma and X-ray signals detected by \textit{GRINTA} we adopted the approach described in Ref.\citenum{Ronchini2022}, and recently refined\cite{DeSantis2026}. This method assumes that the majority of short GRBs are produced by BNS mergers and takes into account the most up to date estimates of BNS local rates derived by the LVK collaboration. To characterise possible scenarios at the time \textit{GRINTA} will operate, we consider
three different networks of GW observatories:
\begin{itemize}[noitemsep, leftmargin=*]
      \item Post O5: 3A$^{\#}$ (two located in the current position of the LIGO antennas in Livingston and Hanford, and the third one in India). This is a realistic scenario expected for the time of the launch of \textit{GRINTA}.
    \item ET in the configuration of 2L, located one in Sardinia and the other in  Euregio Meuse-Rhine site.
    \item ET in the configuration of 2L + CE (40 km arm lenght) in US.
\end{itemize}
Assuming a duty cycle of $85\%$ for each interferometer, 
the maximum achievable redshift for BNS mergers for the 3A$^{\#}$ scenario is $z\sim 1$. Well localised GW sources ($\Delta \Omega \lesssim 100$ deg$^2$) will be concentrated at $z\lesssim 0.5$ (see Fig.\ref{z_hist}). With the advent of the next generation of GW detectors, the BNS detections will be distributed over a wide range of redshift (up to z around 3-5) with a sky localisation $\Delta \Omega < 100$ deg$^2$ for the majority of sources detected at $z\lesssim 1$. The inclusion of just another interferometer in the 3G GW network can improve the sky localisation up to a factor 10 \cite{Ronchini2022}. In all these cases \textit{GRINTA} will provide accurate, sub-arcmin localisation to guide ground-based follow-up observations, fully characterise the source, and identify the host galaxy.

For \textit{GRINTA} operating in conjunction with the 3A$^{\#}$ network, it is estimated that among the $\sim65$ short GRBs/year detected by TED, $\sim2$ are expected to have an associated GW signal. The number of joint detections increases to $\sim50$ per year in the ET scenario, and almost the totality with ET+CE. 
Moreover, the capability of \textit{GRINTA} to directly respond to GW triggers from ground is worth being considered. The FOVs of both TED and HXI match very well the localisation capabilities of upcoming and future GW observatories and enable the rapid repointing of the satellite after receiving a trigger. Two limit cases were considered, where the spacecraft responds to GW alerts within 5 min and 1 hr from the merger, respectively. 

In order to get a quantitative estimate for the above, one may consider the afterglow detection by the HXI and model the afterglow emission following the same prescriptions adopted in \cite{Ronchini2022}. Given the large expected amount of GW triggers for next generation observatories, the information given by GW parameter estimation (sky localisation, luminosity distance, inclination angle and related error) is leveraged to restrict the selection of a golden sample to be followed up with HXI. 
As a result of adopting this method, a few extra joint detections are expected in the local Universe: in the case of A$^{\sharp}$, the number of joint detections doubles, bringing the total number of joint detections over the nominal lifetime of GRINTA (two years) to about a dozen. 

\subsection{Particle Acceleration in Extreme Environments}

\subsubsection{Ultra high energy cosmic rays and neutrinos}

The detection of high-energy neutrinos has opened a new window to the high-energy Universe and may provide the long-sought breakthroughs. Combined with observations of hard X-rays provided by \textit{GRINTA}, crucial and complementary information about the most violent phenomena in the Universe will be obtained. High-energy neutrinos and X-rays are directly linked to interactions of UHECR in or around their acceleration sites and provide complementary information about them. Hard X-rays provide detailed information about non-thermal processes, magnetic fields, as well as emission and acceleration mechanisms. Neutrinos on the other hand are weakly interacting particles and are direct tracers of hadronic interactions. They can escape the densest astrophysical sources and travel long distances from the source. Their detection is difficult and to date, only a small number of individual high-energy neutrino events could be tentatively associated with multi-wavelength counterparts.
The most prominent is the blazar TXS~0506+056 that could be linked at the 3 $\sigma$ level to the high-energy neutrino IceCube-170922A during a several weeks long flaring period in 2017~\cite{science2018}. 

Recent analyses provide first indications for associations between high-energy neutrinos and X-ray Bright Seyfert Galaxies~\cite{https://doi.org/10.48550/arxiv.2602.10208}. Observations in the hard X-ray domain as provided by \textit{GRINTA} are the most promising avenue to resolve the UHECR/high-energy neutrino puzzle: the large FoV of \textit{GRINTA} allows to cover the entirety of the localisation uncertainty region of individual high-energy neutrino events and the covered energy range is the most sensitive one to telling different emission scenarios apart. 
The timescale of \textit{GRINTA} will overlap with the operation of the next generation neutrino telescopes IceCube-Gen2 and KM3NeT and thus create exciting synergies for joint multi-messenger studies.

\subsubsection{Highly magnetised neutron stars}
Magnetars and soft $\gamma$-ray pulsars are important sites of particle acceleration that are known to give rise to high energy phenomena and thus, are sought to be studied in hard X-rays. 
With a magnetic field as high as $\sim 10^{14}-10^{15}$~G,  
larger than the limit for quantum effects to be important,
magnetars are the most strongly magnetised neutron stars \cite{mereghetti2015,kaspi2017}. Their X-ray/gamma-ray emission is characterised by different variability phenomena, including short bursts ($L \sim 10^{39}-10^{42} \rm \, erg \, s^{-1}$), intermediate flares ($L\sim 10^{42}-10^{44} \rm \, erg \, s^{-1}$) and  spectacular giant flares reaching peak luminosities up to  $L \sim 10^{47} \rm \, erg \, s^{-1}$. 
Most magnetars are transients: they spend most of the time at quiescent X-ray luminosities $\sim 10^{33}-10^{34} \rm \, erg \, s^{-1}$, and occasionally experience weeks to months-long outbursts during which the emission increases by \(1-2\) orders of magnitude.  This implies a Galactic population much larger than the currently known sample of about 30 sources. 
The cause of the magnetar activity, and the triggering mechanism of the instability that causes their extreme variability are still poorly known \cite{turolla2015}. 
Due to its large field of view and energy range, TED will detect magnetars short bursts down to fluences at least one order of magnitude below the current limit.
Magnetars emit almost half their luminosity in the form of persistent, hard,  non-thermal components of  magnetospheric origin, extending to at least 100~keV with no observed cutoffs. 
High resolution, phase-resolved spectroscopy of these hard tails with 
\textit{GRINTA} will allow us to probe the unknown magnetar field structure, and  to finally understand the acceleration mechanisms of magnetospheric particles under such extreme conditions.  

In addition, the systematic study of the high energy activity and the detection of new candidates by \textit{GRINTA} will be key in shedding light on the connection between magnetars and other classes of sources, such as the long and short GRBs, the mysterious Fast Radio Bursts, as well on the magnetar role as  possible sources of neutrinos, very high energy cosmic rays, and gravitational waves \cite{Arons2003,zhang2020,DallOsso2021}.

\textit{GRINTA} will also contribute significantly to the study of the population of soft $\gamma$-ray pulsars. These spin-down powered sources are generally younger and 
more energetic than the \textit{Fermi} LAT pulsar population, and the members are often associated with TeV counterparts. Moreover, the majority lacks clear radio- and $>100$ MeV pulsations. The 20 keV\(-\)10 MeV band, well covered by \textit{GRINTA}'s HXI and TED, contains crucial information about the nature of the non-thermal emission processes taking place in the pulsar's magnetospheres. 
\subsubsection{Fast radio bursts}
Fast radio bursts (FRBs) are $\sim$ms long radio flashes of extragalactic origin \cite{2023RvMP...95c5005Z}, now detected at rates of several events per day \cite{Cordes2019}. Their origin remains uncertain, with models invoking magnetar flares, relativistic shocks, and compact object mergers. Therefore,
even the detection of a single well-characterised event in the X-rays would provide key constraints on the emission mechanism, allowing to discriminate between magnetospheric and shock-driven scenarios through its spectral properties and the relative timing between the radio and high-energy signals.
In this context, \textit{GRINTA} will provide a decisive advance through continuous wide-field monitoring in the hard X-ray band (5--200 keV), enabling systematic searches for prompt high-energy counterparts. Assuming a radio-to-X-ray fluence ratio comparable to SGR~1935+2154 ($\eta \sim 10^{-5}$), it will be sensitive to Galactic-like events across the Local Group, and to more energetic bursts out to a few Mpc.
These observations will directly probe the prompt emission mechanism and energetics of FRBs, testing magnetar-based scenarios and constraining the distribution of the radio-to-X-ray fluence ratio. 


In synergy with next-generation radio facilities such as SKAO, CHORD, and DSA-2000, which will deliver large samples of well-localised FRBs, \textit{GRINTA} will provide the crucial high-energy counterpart information needed to link prompt emission to the properties of the central engine.

\subsection{Accretion Flows and Collimated Outflows in Compact Objects}

\subsubsection{Accretion disk, corona, and jet connection in transient X-ray binaries}


Outbursts, flares, and state transitions of X-ray binaries are frequent and can occur at any time. The wide FoV of \textit{GRINTA}’s instruments will be an asset for serendipitously detecting early (hard) stages of outbursts, a phase still poorly studied.  
\textit{GRINTA}’s spectral capabilities will then be an asset to study the fine physics, interplay, and correlations of the high-energy components from 5 keV to 200 keV, and their evolution/causal relation at state transitions.
GRINTA will provide early-warning
signals for the onset of X-ray binaries outbursts, supporting many follow-up campaigns including e.g. in the optical/IR and radio. 
Simultaneous fast photometry can distinguish between both components (optical emission due to reprocessing should be a lagged and smeared version of the X-ray emission).  For example, \textit{INTEGRAL} observations of V404~Cygni revealed optical/X-ray correlations, probing different underlying physical mechanisms \cite{2015A&A...581L...9R,Alfonso-Garzon2018}.
In HMXBs, emission from the donor star usually dominates at these wavelengths. However, Be/XRB models predict a transitional accretion disk could form, and reprocessing by this disk and/or the Be circumstellar disk could be explored by measuring the lags between X-rays and IR/optical/UV emission. 

By 2030, the Extremely Large Telescope (ELT) will be operative and will be able to provide accurate optical and NIR spectra, and precise NIR and MIR photometry, ideal for studying XRBs in crowded regions such as the Galactic Centre. 
Simultaneous radio observations will probe the relativistic jets and their connection with the accretion
flow in BH XBs and some types of NS XBs. In the 2030s, the Square Kilometer Array Observatory (SKAO) is expected
to be fully operational, while the next-generation VLA is planned to come online halfway through the
decade.
\subsubsection{Origin of high-energy emission in blazars}

Blazars, comprising flat-spectrum radio quasars and BL Lac objects
(further classified into LBL, IBL, and HBL), are the most powerful andvariable Active Galactic Nuclei (AGN). They feature relativistic plasma jets oriented within a few degrees of the line of sight. Their spectral energy distribution (SED) exhibits a characteristic double-hump profile. The low-energy peak is widely attributed to synchrotron radiation from relativistic electrons spiralling in the jet's magnetic field, whereas
the origin of the high-energy peak remains controversial \cite{Ghisellini2017}. Leptonic
models attribute this high-energy emission to Inverse Compton (IC) or
Synchrotron Self-Compton (SSC) scattering, whereas alternative hadronic
scenarios invoke processes such as proton synchrotron radiation.
\textit{GRINTA} will help distinguish these scenarios by monitoring the critical SED transition region between the two emission components.
Simultaneous, co-spatial high-energy neutrino detections will be crucial for constraining jet composition, guiding SED modeling, and evaluating
blazars as potential primary sources of ultra-high-energy cosmic rays
(UHECRs). GRINTA's wide field-of-view (FOV) ensures frequent hard X-raysky coverage and rapid responses to blazar fl ares. This guarantees that outbursts, particularly from neutrino-source candidates disseminated by IceCube and KM3NeT~\cite{science2018}, are fully resolved.

Among blazars, BL Lac objects exhibit lower luminosities and less efficient cooling, shifting both SED peaks to higher frequencies. The extreme high-peaked BL Lac sub-class (EHBL)\cite{Foffano2019, Acciari2020} features extreme properties and intense emission states well-suited for detection by \textit{GRINTA} in coordination with current and future high-energy facilities. Concurrently, the joint response of \textit{GRINTA} and the upcoming Cherenkov Telescope Array Observatory (\textit{CTAO}—fully operational during \textit{GRINTA}'s lifetime—will map the multi-wavelength evolution of flare parameters with unprecedented temporal resolution.

\subsubsection{Systematic studies of tidal disruption events}
Wide-field surveys have recently enabled the discovery of large samples of Tidal Disruption Events, observed as luminous flares from otherwise quiescent galactic nuclei (see e.g. \cite{komossa2015,jonker2021}). While most TDEs are dominated by thermal emission from the accretion flow, a small fraction shows non-thermal components and, in rare cases, launches relativistic jets.
Despite this progress, the origin of the high-energy emission and the conditions leading to jet formation remain poorly understood. In particular, the hard X-ray band provides a key diagnostic to distinguish between coronal emission, obscured accretion, and jet-related processes.
In this context, \textit{GRINTA} will play a crucial role by providing sensitive coverage above 5~keV, enabling systematic follow-up of TDEs discovered by wide-field soft X-ray and optical surveys and testing 
the presence of relativistic jets in synergy with future radio facilities (e.g. SKAO).

\subsection{Other Science Opportunities }
Beyond short-lived transients, the hard X-ray sky contains faint, persistent emitters often undetected in single observations. These sources can be unveiled via survey science by stacking multiple observations.

Current catalogs of the $\sim2000$ hard X-ray sources are based largely on \textit{INTEGRAL} and \textit{Swift} surveys, which are dominated by Galactic binaries and
extragalactic AGN, along with several emerging populations. The broader energy range of \textit{GRINTA}/HXI will advance the study of all these
classes. Key emerging Galactic populations for \textit{GRINTA} to investigate include: (a) obscured binaries, which revolutionised the understanding
of Galactic supergiant binaries by revealing highly obscured persistent emitters undetected at other wavelengths; (b) supergiant fast X-ray
transients, which reveal short-term accretion regulated by a complex interplay of stellar-wind density, geometry, and neutron star magnetospheric regimes; and (c) accreting white dwarfs, a local population established as common X-ray emitters by \textit{INTEGRAL} and \textit{Swift}. This population comprises magnetic cataclysmic variables (CVs), predominantly intermediate polars (IPs) that constitute ~6\% of \textit{INTEGRAL}
sources, alongside a small number of symbiotic and nova-like systems.
Crucially, \textit{GRINTA}/HXI will provide a complete, flux-limited hard X-ray sample of short-period IPs, which are expected to have a space density more than an order of magnitude higher than their brighter, long-period
counterparts\cite{Pretorius2014}. 

\section{Science instruments}
The payload concept is based on the latest developments in CdTe detector technology and in compact scintillator devices with Silicon Photomultiplier (SiPM) readout. CdTe devices have been already flown on board \textit{INTEGRAL} \cite{Wink2003A&A...411L...1W}, Solar Orbiter \cite{Krucker2020}, and more recently on \textit{SVOM} \cite{Shunjing2020}. In particular, the ISGRI detector \cite{Lebrun2003} was operational for 22 years in space. 

The performance requirements for the two instruments on board \textit{GRINTA} are listed in Table~\ref{tab:perf-reqs}. An array of $16\times16$~CdTe pixel detectors will be the core of the Hard X-ray Imager (HXI) featuring a spatial resolution of 1mm, thus allowing for high resolution mapping of the sky by decoding with the coded mask, reaching a source location accuracy of $\sim40$~arcsec.
For bright transients and GRB detection a set of 16 scintillator detector modules, composing the Transient Event Detector (TED) with size 150~cm\(^2\) each will be used to cover $\sim8$~sr of the sky. This instrument will have a localisation capability that is well matched to the HXI's FoV for the purpose of transient's follow-up.

\subsection{Transient Event Detector}
\subsubsection{Design and operating principle}

\begin{figure}[b]
  \centering
  \includegraphics[height=8cm]{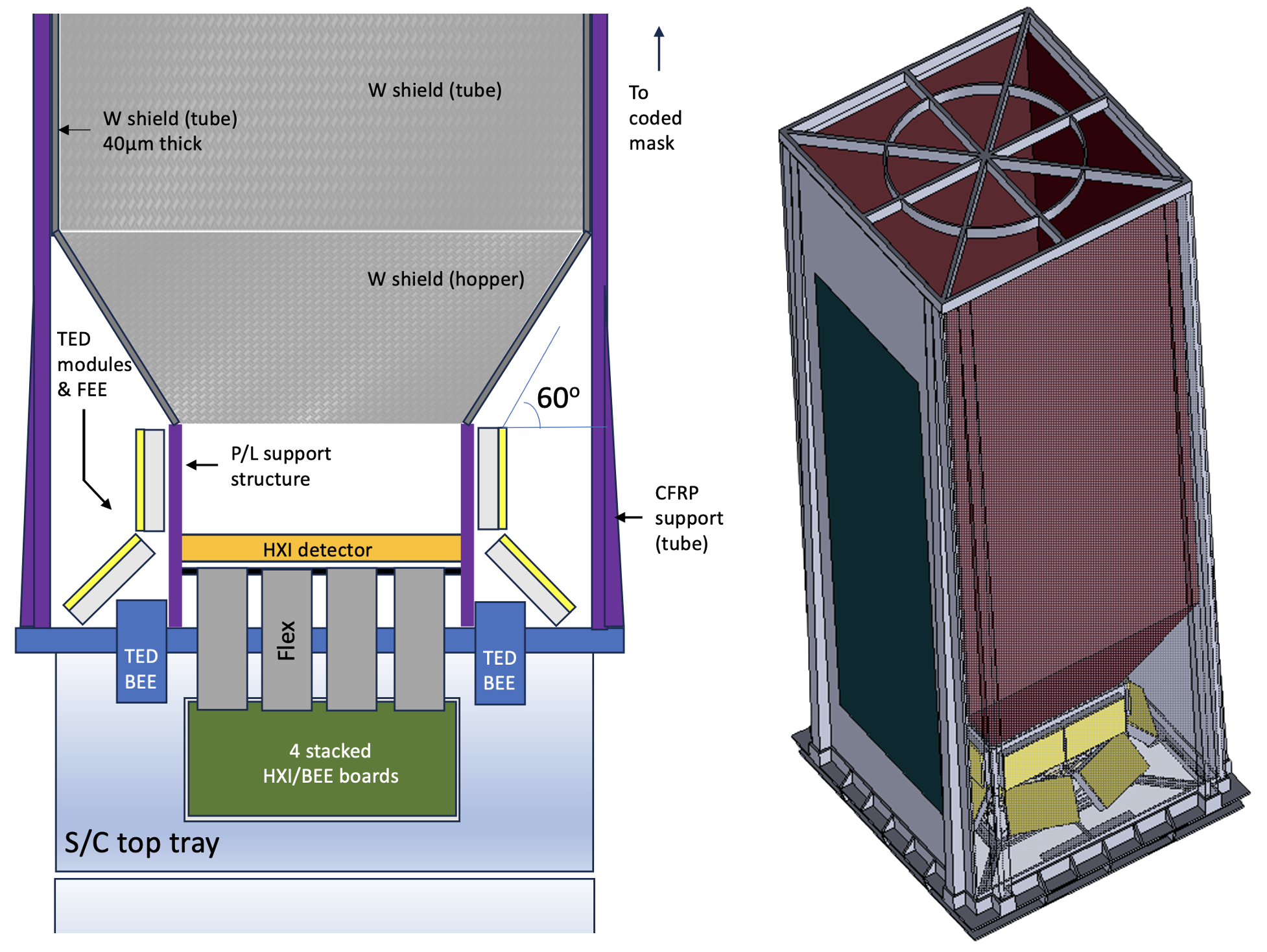}
    \caption{(Left) Schematic of the \textit{GRINTA} detector payload elements (side view). (Right) 3D- view of payload accommodation. The passive shield system is shown in dark brown. On the top, the coded mask supporting structure.}
  \label{fig:grinta_layout}
\end{figure}

The TED instrument will be based on a set of 16 detector modules, each one hosting 24 scintillators tiles having a thickness of 8\,mm, and their associated FEE. TED will use Gadolinium Aluminium Gallium Garnet (GAGG) crystals optically coupled to Silicon photomultiplier (SiPM) readout arrays to monitor the soft gamma-ray sky with a 10-\(\mu\)s time resolution.  
 Each module sizes 150~cm\(^2\) and is positioned around the payload support structure as shown in Fig.~\ref{fig:grinta_layout}. Eight modules will be placed on the sides of the HXI detection plane: they will also act as active shields, being oriented 
with their main axis at 90\(^{\circ}\) from the HXI pointing axis. The remaining eight modules will be placed around the PLM base with an inclination  
of 45\(^{\circ}\) from the HXI pointing axis.

\begin{figure}[ht]
  \centering
  \includegraphics[height=5cm]{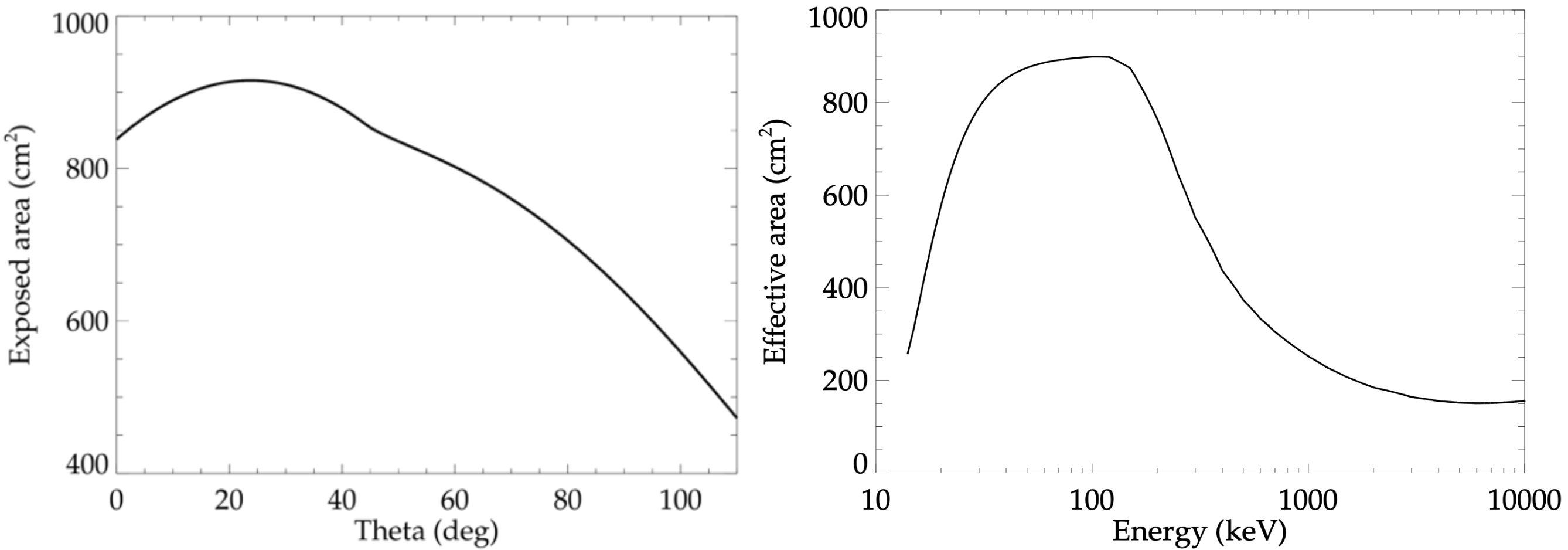}
  \caption{(Left): TED exposed detector area as a function of polar angle from the S/C boresight. (Right): Effective area of TED at 30 degrees polar angle.}
  \label{fig:ted_exp_area}
\end{figure}

\begin{table}[ht]
\caption{\textit{GRINTA} performance requirements. }
\label{tab:perf-reqs}
\begin{center}    
\begin{tabular}{ |p{3.5cm} |p{6.9cm} |p{4.3cm}|}
\hline
\textbf{Parameter} & \textbf{HXI} & \textbf{TED}  \\\hline                    
  Energy Range         &  5--200\,keV  & 20\,keV--10\,MeV    \\\hline
  Spectral Resolution (FWHM)  &  1\,keV@60\,keV  & \(\sim\)25\%\,@60\,keV \(\sim\)10\%\,@500\,keV     \\\hline
  Field of View        &   $\sim400$\,sq.deg with $>50$\% sensitivity & $\sim8$ sr    \\\hline
  Angular Resolution   &   3.8' & N/A     \\\hline
  Time resolution & $<50$~
  $\mu$s  & $<20\mu$s \\\hline
  Source location accuracy\,(8$\sigma$) & {45} $\!\!^{\prime\prime}$ & 80\% of GRBs within $10^{\circ}$  \\\hline
  Effective area       &  $\approx400$cm$^2$ @20 keV & \hspace{-2.5mm}  $\approx\,900$cm$^2$\,@100 keV  \\\hline
  Sensitivity  (SNR=3, 10\(^4\)\,s) & 5--30\,keV: 1.2\(\times\)10\(^{-3}\)\,ph/cm\(^2\)/s [1.0 mCrab]
  30--60\,keV: 6.9\(\times\)10\(^{-4}\)\,ph/cm\(^2\)/s [6.0 mCrab] 
  60--120\,keV 5.1\(\times\)10\(^{-4}\)\,ph/cm\(^2\)/s [9.5 mCrab] & 
  \(<\)0.5\,ph/cm\(^2\)/s, 50--300\,keV  \\\hline
\end{tabular}
\end{center}
\end{table}
The latter modules will be passively shielded on the side opposite respect to the scintillator,  by a 1~mm thick tungsten sheet in order to prevent most of the albedo gamma-rays from reaching the active detector volume. 
This spatial configuration is designed to maximise the detector area exposed over a wide range of off-axis angles relative to the HXI pointing direction, thereby maintaining high sensitivity even to events occurring far from the S/C boresight (see Fig.~\ref{fig:ted_exp_area}, left panel). TED's localisation capability ($\sim10^{\circ}$) is well matched to the HXI's FoV for the purpose of transient's follow-up.  The total detection area of the instrument is 2400\,cm\(^2\), resulting in an effective area of $\approx900$~cm which will allow a better (factor $\sim1.7$) sensitivity than the current operational \textit{Fermi}/GBM \cite{Meeg2009ApJ...702..791M}. 

\begin{figure}[b]
  \centering
  \includegraphics[height=4cm]{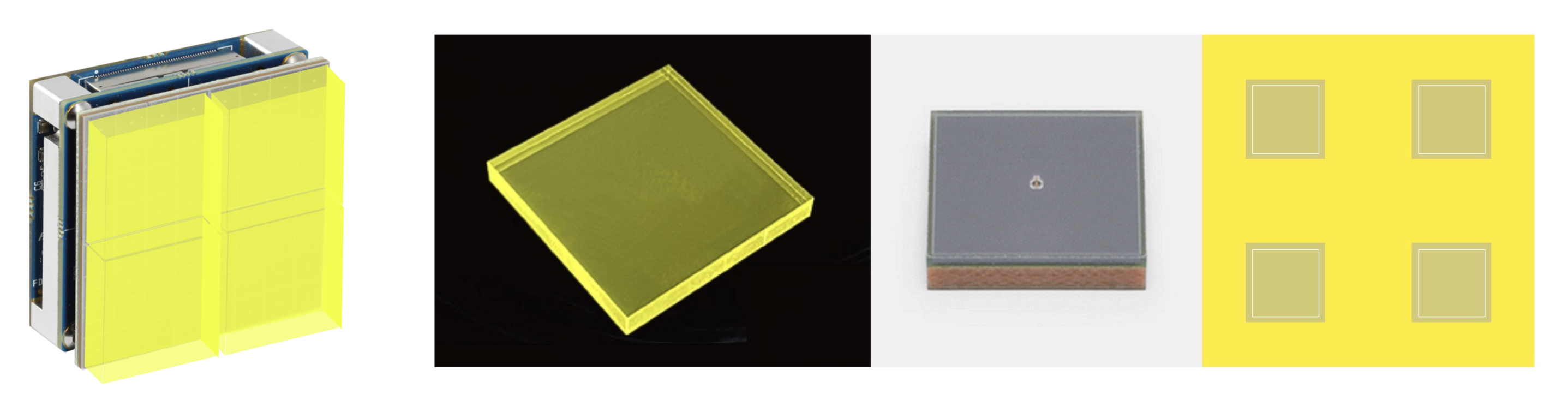}
  \caption{(Left)Schematic of a TED detection unit with 2\(\times\)2 scintillator tyles and readout system. Each GAGG scintillator size is 2.5\(\times\)2.5cm$^2$. The readout of the entire unit is handled by a single 16-channel ASIC. (Right) photograph of a GAGG scintillator tile of size 2.5\(\times\)2.5cm$^2$ (credit:Epic Crystal). Centre: the Hamamatsu MPPC S14160-6050HS, used for the readout of the scintillation light, with 6x6 mm$^2$ photosensitive area. (Right) Schematic of a single tile configuration with 4 MPPCs. }
  \label{fig:ted_scheme}
\end{figure}

TED will use GAGG scintillators viewed by SiPM arrays (see Fig.\ref{fig:ted_scheme}). The SiPM readout technology, widely used in health diagnostic applications has already flown on several astronomical satellites (e.g. \textit{GECAM}\cite{Qiao2024}, \textit{GRBalpha}\cite{Pl2020}, \textit{EIRSAT}\cite{Murphy2022}). 
As demonstrated with earlier missions (i.e. \textit{Fermi}/GBM and \textit{CGRO}/BATSE), the count rate for uncollimated detectors varies smoothly in LEO. Thus short, impulsive events will appear super-imposed on the background count-rate of the detector.  
By monitoring detector count rates with onboard ratemeters, it becomes possible to detect such events (i.e., GRBs) by identifying statistically significant increases above the background level across multiple detectors, on timescales ranging from milliseconds to a few seconds.

Once an event is detected, the onboard software determines its location by analysing the relative count rates—after background subtraction—across all 16 modules.  These rates will be compared to tables of expected relative count rates for a grid of locations in spacecraft coordinates.  The location of the event is then converted to Right Ascension and Declination using the spacecraft pointing information and processed by the \textit{GRINTA} Data Processing Unit (DPU) for possible re-pointing with the HXI.  It is expected to localise $\approx80$\% of detected events to within \(10^{\circ}\). 

The count-rate data and the event location will be transmitted to ground and shared via Gamma-ray Coordinates Network (GCN) to the community for potential multi-wavelength follow-up.  

\subsubsection{Implementation scheme}
In the baseline configuration, one TED module will be composed of  \( 3 \times 2\) detection units each one hosting an array of four  \(2.5 \times 2.5 \times 0.8 \) cm\(^3\) GaGG crystals, yielding a total detection area of 150~cm\(^2\), see Fig.~\ref{fig:ted_scheme}, left panel. The on-board data implementation scheme is such that the 16 modules are organised in 4 ``walls'',
two for the lower sector and another two for the upper sector. Each wall will then  consist of 4 modules and will be interfaced to a Back-end Electronics (BEE) board. 

The detectors will be within a protective structure, which shields them from visible light but remains transparent for gamma-rays in the part facing the mask. This housing can be coated with reflective material to reduce the heating from environmental radiation according to the approach of dimensioning a passive thermal control that tends to keep the various detectors cold.
For the TED modules the operating temperature will be kept between -10$^{\circ}\text{C}$ and 
10$^{\circ}\text{C}$, being stable at the 1-2 degrees level during a typical observation 
($\sim10$ks). 

\subsubsection{Frontend electronics}
As baseline, each scintillation tile will be optically coupled to 4 SiPMs yielding a total of 16 channels to be readout from a single detection unit (Fig.~\ref{fig:ted_scheme}, right panel). The four SiPMs will be mounted on a single PCB to interface the scintillator's surface. The SIPHRA ASIC developed by IDEAS \cite{stein2019} is the preferred frontend solution due to its low power consumption (21mW for the CMIS version) and high component TRL. The ASIC can be purchased already wire-bonded in the IDE3380 card provided by IDEAS. 
A specific Module I/F card (MIC) hosting 6 SIPHRA ASICs will be designed with connectors mating with the front-end. This includes input signals (Ain1 – 16) for six ASICs (U1 – U6) and detector bias voltages (DetBias1 – 6). This card element has a well established design heritage from the IDEAS/ROSSPAD module design. 

\subsubsection{Backend electronics}
\begin{figure}[t]
  \centering
  \includegraphics[height=7cm]{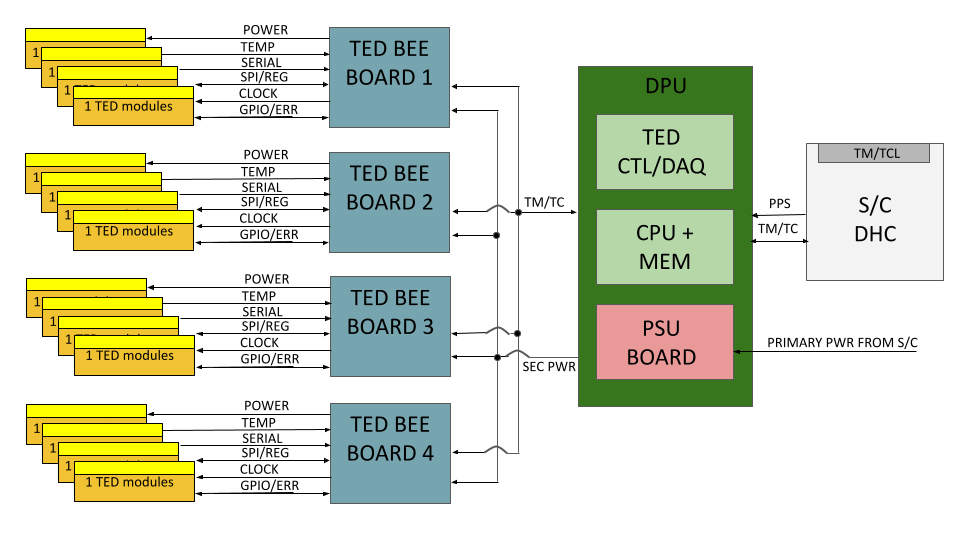}
  \caption{TED Backend Electronics Block Diagram.}
  \label{fig:TED_scheme-BEE.pdf}
\end{figure}

There will be four BEE boards serving each one a TED``wall". 
They will manage the read, data buffering to the DPU, management of power supplies, protections, and fail isolation for each TED module. 
In the present version the BEE is composed of 4 Field Programmable Gate Array (FPGA) boards that will interface the DPU on one side, and the MICs connected to the 4 modules assigned to each board (see Fig.\ref{fig:TED_scheme-BEE.pdf}). The interface between the modules and the Back End Electronics 
includes the following functions:
\begin{itemize}[noitemsep, leftmargin=*]
    \item Power: Supplies positive voltage to the SIPHRAs and negative voltage to the SIPMs in CMIS mode.
    \item Temperature: Monitors up to four temperature sensors per module.
    \item Serial Data: A 2.5MHz interface for data retrieval from the SIPHRAs.
    \item SPI: Used for SIPHRA register configuration and status monitoring.
    \item Clock \& GPIO: Provides the system clock and a multi-purpose pin for error handling or external triggers.
\end{itemize}

The BEE boards manage component powering and biasing, data buffering for the DPU, and failure isolation to protect the sensor integrity. Communication between the BEEs and the \textit{GRINTA} DPU is handled via TM/TC and power buses. Finally, the DPU executes high-level tasks such as raw data analysis, system state management, and power distribution across all subsystems. As a requirement, each TED module will be autonomous.  Thus a failure of a single module can be isolated, leaving the functionality of the others intact.  


\subsection{Hard X-ray Imager}
\subsubsection{Design and operating principle}

The Hard X-ray Imager is a coded-mask telescope. This imaging technique implemented in \textit{Granat} (1989), \textit{INTEGRAL} (2002), \textit{Swift} (2004) and \textit{SVOM} (2024) allows localising astrophysical sources in the hard X-ray range where focusing optics become inefficient. The main elements of this instrumental concept is the coded-mask and the detection plane. The design drivers are the followings.

\begin{itemize}[noitemsep]
    \item The imaging performance (angular resolution, field of view) is defined by the mask geometry, detector size and the distance from the mask to the detection plane
    \item The spectral performance (energy range, spectral resolution) is defined by the detector choice including the front-end electronics.
    \item The instrument sensitivity is defined mainly by the detection plane surface and the X-ray background rejection capability. 
\end{itemize}

For the imaging performance, the theoretical point source location accuracy (PSLA) of a coded mask telescope, expressed as 90\% c.l. position error is a function of the detector pixel size, d, the mask-detector distance, H, the resolution parameter r (mask to pixel linear size) and the signal-to-noise ratio \cite{goldwurm2022}: \\ $PSLA(SNR) \approx \arctan\left(\frac{\sqrt{ln(10)}}{SNR}) \cdot \frac{d}{H} \cdot \sqrt{r - \frac{1}{3}}\right)$. \\

\begin{figure}[b]
  \centering
  \includegraphics[height=8.5cm]{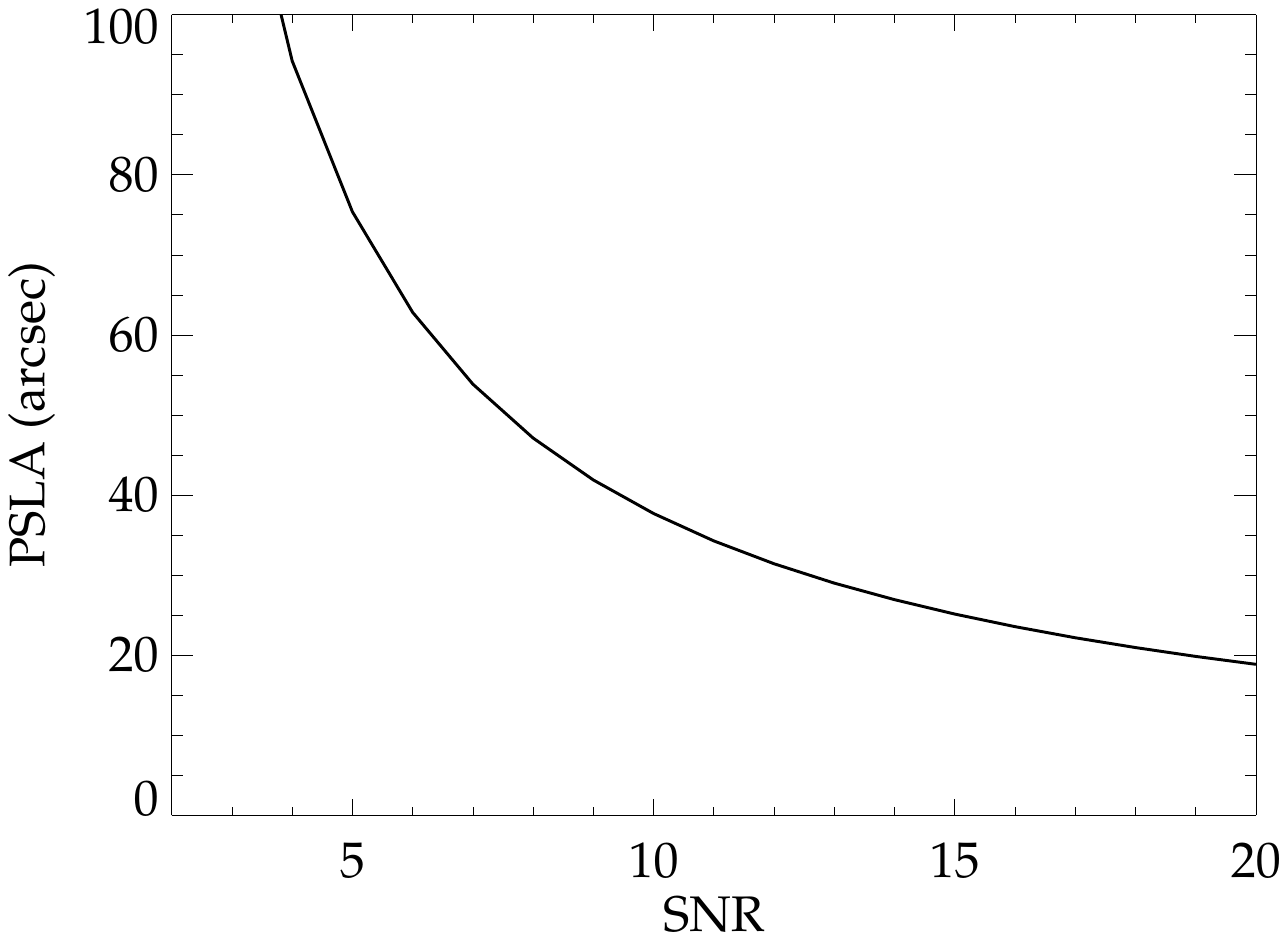}
  \vspace{-5mm}
  \caption{HXI Point source location error (90\% c.l) as a function of the signal-to-noise ratio.}
\label{fig:psle}
\end{figure} 

For the \textit{GRINTA} baseline parameters (r=2.0, H=180cm, d=1mm) the curve in Fig.~\ref{fig:psle} represents the theoretical PSLA as a function of the SNR. For sources close to the detection limit ($\approx5\sigma$) the accuracy is $\sim1'$, whereas for strong signals the PSLA can even approach 20". In reality, systematic effects in real applications can affect the location error up to a factor $\sim1.5$, as was the case for \textit{INTEGRAL}/IBIS. These eventually include the finite attitude accuracy but also instrumental biases. However, in the worst case, expectations for \textit{GRINTA} are to keep a $\sim1$ arcmin localisation performance for on-board automated detections.

\subsubsection{Implementation scheme}

\textbf{Coded Mask}- The HXI coded mask is made of tungsten square blocks disposed within a surface area of 3560 cm\(^2\) with an open fraction \(= 1/2\) as used by \textit{Swift}/BAT \cite{Barthelmy2005}. 
The mask is separated from the detector by 1.8 m yielding a 10\(^{\circ}\)x10\(^{\circ}\) fully-coded FoV (\(29^{\circ} \times 29^{\circ}\) at zero response).
The supporting carbon-fiber telescope tube includes a $40\mu\text{m}$ tungsten shield to attenuate out-of-field X-ray photons.

\noindent\textbf{Hard X-ray Detector (HXD)}-
The HXD utilises a pixelated Cadmium
Telluride (CdTe) semiconductor array, offering superior hard X-ray quantum efficiency at moderate cooling temperatures relative to silicon
or germanium. The readout chain comprises: (i) the CdTe crystals and analog front-end ASICs cooled to \(-20^{\circ}\)C (cold front-end electronics,
CFEE); (ii) a warm front-end electronics (WFEE) unit for voltage regulation and analog-to-digital conversion; and (iii) a digital back-end electronics (BEE) unit.

The detection plane comprises a \(16 \times 16\) array of CdTe crystals. The baseline configuration employs 1.5~mm thick Schottky Al pixel detectors, selected for low leakage current, high manufacturing heritage, and
favorable cost and lead-time constraints. Each crystal features a segmented anode forming a \(16 \times 16\) pixel array enclosed by a guard ring. The pixels are 1 mm wide, with a 1 mm inter-crystal spacing. The total detection area is 824~cm\(^2\), providing
655~cm\(^2\)cm² of active area after accounting for an 8\% dead space.

\begin{figure}{h}
    \centering
    \includegraphics[height=7.5cm]{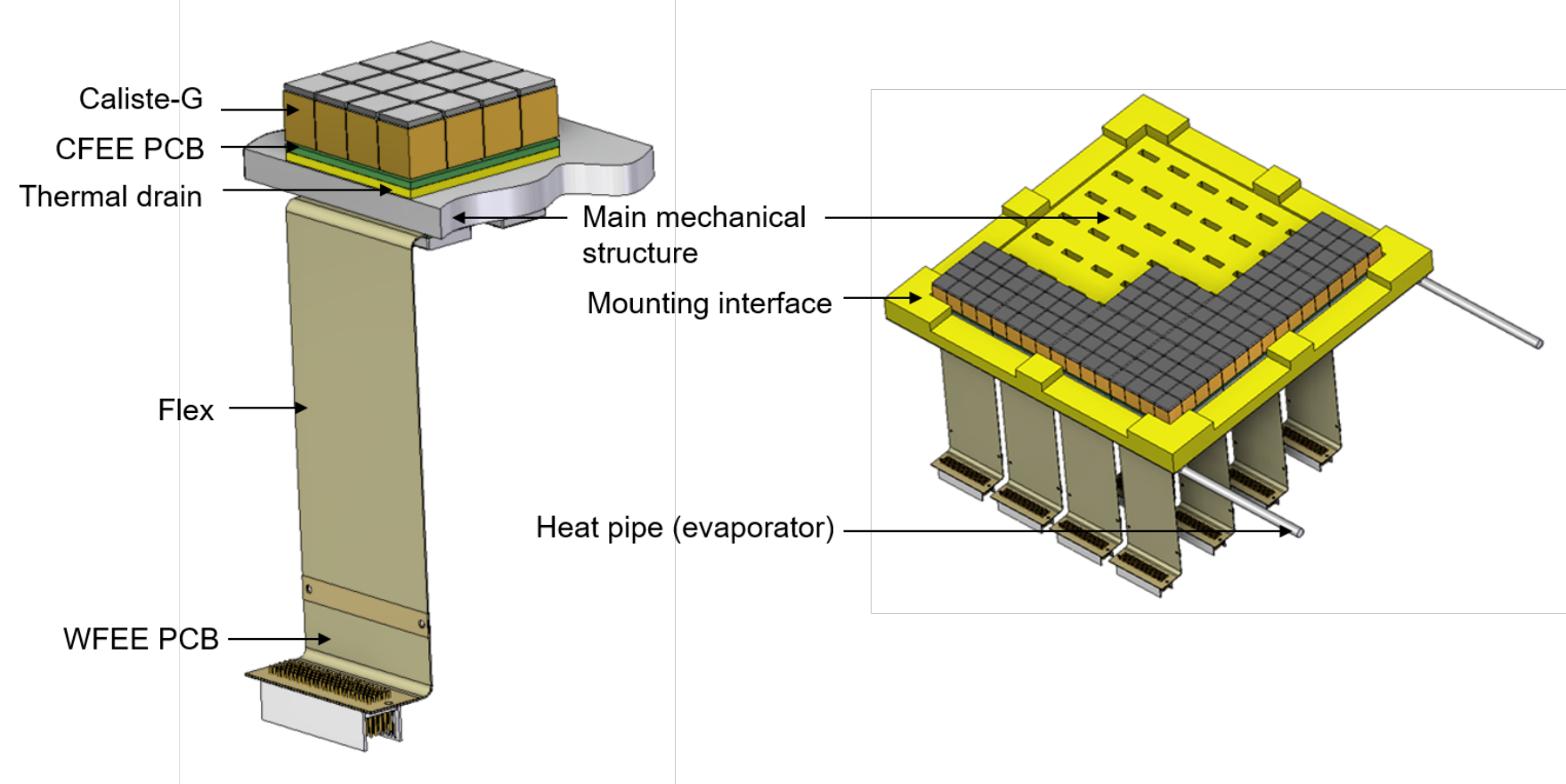}
    \caption{(Left) one detection unit of the HXD holding 16 Caliste-G modules. (Right) detection plane of the Hard X-ray imager.}
    \label{hxi_design}
    \vspace{-4.5mm}
\end{figure}

To interconnect the 65.536 spectroscopic channels, the front-end ASICs are oriented perpendicular to the detection surface and stacked using 3D~PLUS company packaging technology. This approach ensures a highly compact, modular design with excellent spectroscopic performance. This hybrid architecture, designated Caliste, has been developed since 2007 and is successfully flight-proven on the Solar Orbiter/STIX instrument.\cite{Limousin16}. The
\textit{GRINTA} variant, Caliste-G, incorporates eight IDeF-X HD ASICs to manage 256 spectroscopic channels alongside a global trigger logic\cite{Gevin21}.
Structurally, the detection plane is divided into four autonomous sectors. Each sector consists of a row of four detection units, with each unit housing a \(4 \times 4\) array of Caliste-G modules. Independent BEE
units manage data acquisition and control interfaces for each sector.
Following assembly and performance verification, these units will be integrated into a primary structure fabricated from aluminum alloy or AlBemet (see Fig. \ref{hxi_design}). 
Within this structure, two heat pipes and two
thermal spreaders will dissipate power from the Caliste modules to a $1.5 \rm \, m^2$ radiator. Flight heritage and thermal models confirm this configuration can maintain a continuous CdTe operating temperature of \(-20^{\circ}\)C in a 600~km circular orbit.

\subsubsection{Back-end electronics}
The back-end electronics (BEE) of the HXD will be responsible for operation and readout of the detection plane. It will also host housekeeping circuitry for low and high voltage supplies. The organisation and layout of the BEE follows the sectoring concept of the HXD: Each BEE sector will be completely self-contained and independent of the others and will serve the corresponding quadrant of the HXD. In this way, propagation of a failure to another quadrant is avoided and a graceful degradation of the performance of the instrument is achieved in such a situation, as the continuous operation of the rest of the instrument can be ensured.

Each BEE sector will perform its control, readout and operation tasks fully in the digital domain using an RTG4 FPGA. The radiation hard, low power consumption RTG4 by Microsemi/Microchip offers the high number of I/O pins required to connect to the WFEE. In addition, it allows to store configuration values for the front-end ASICs as well as calibration tables internally, avoiding additional external memory devices. Precise timing and real-time (but simple) data pre-processing also require a highly parallel (one for each Caliste unit) pipeline concept that is traditionally best implemented in an FPGA as opposed to a microcontroller chip. Each BEE sector has an output buffer and its own TM/TC, HK and power interface to the main data processing unit. For the selected read-out configuration, a number of external LVDS drivers is necessary as extensive multiplexing of signals is being applied in order to keep the interface between BEE and detection plane feasible.

After the event-triggered data acquisition, a precise timestamp will be attached to each event and the data move forward through the pre-processing pipeline. Automatic offset, noise, common-mode and gain corrections will take place in this pipeline such that no additional dead time (besides the read-out time) is inflicted by the processing.

\subsubsection{Operation principle and background rejection}
As soon as a photon is detected in the detection plan, the associated Caliste module sends a trigger signal to the BEE FPGA that starts the readout. The full Caliste detector unit is frozen during the 20~µs readout and then again operational for new photon detection. The detector units are operated independently and can be processed in parallel. Photon lists with pixel coordinates and raw amplitudes are sent to the DPU.

Based on the background rate of \textit{INTEGRAL}/ISGRI, one can expect a background rate of \( \sim 100\) cts/s for the HXD.  
Joint detection of the same event by HXI and TED could be used for either coincidence or anticoincidence. High-energy particles will be removed by anticoincidence, whereas hard X-ray and gamma ray photons may be used for Compton imaging. Indeed, it can be envisaged that events seen by one TED module and HXI, with total energy in the sub-MeV and MeV range, may be a photon whereas events seen in coincidence with several TED modules are more likely to be particles. This logic will be studied in more details during the first phase of the project.
\color{black}

The HXI timing resolution will be $<$50~$\mu$s.  With this time resolution the expected transmission rate is $\sim4$~Gbit/day.


\subsection{\textit{GRINTA} Data Processing Unit}

\subsubsection{DPU functions}

The \textit{GRINTA} DPU manages payload instrument control, data processing, and power distribution to the instruments. The DPU is housed in the upper sector of the S/C, close to the BEE boards of both instruments. The DPU executes the following core functions:
\begin{itemize}[noitemsep, leftmargin=*]
\item{control instrument operational modes, timing, and synchronisation}; 
\item{process scientific data, including trigger generation, event filtering, image analysis, and source localisation for TED and HXI;}
\item{handle scientific and HK data streams to/from the BEE units;}
\item{compress, packetise, and store scientific and HK data;}
\item{transmit data to the S/C OBDH system for storage and downlink;}
\item{generate alert messages (e.g., GCN format), control the P/L alert transponder, and route uplink/downlink messages via the ISL;}
\item{issue autonomous spacecraft repointing requests;}
\item{generate and distribute secondary power to the BEEs.}

\end{itemize}

\subsubsection{DPU architecture }
The baseline system for the \textit{GRINTA} DPU is identified as the GR740 provided by FrontGrade Gaisler, 
based on the quad-core, fault-tolerant LEON4FT SPARC V8 processor. This microprocessor 
can easily handle PUS (Packet Utilization Standard) packetisation and compression for the 
volume of data generated by GRINTA, whilst keeping power consumption low. The GR740 has 
been extensively qualified for rad-hardness and has been proven in-flight on different satellites and also 
implemented to assist operations in rovers for planetary exploration. 

The DPU design consists of a number of PCB frames which can be modularly stacked and 
inserted directly in the P/L upper tray (no need for a mechanical box). One single board will host the GR740 processor plus a RTG-4 radiation-tolerant FPGA, that is used to cover a number of H/W support functions. To reach the required level of
timing accuracy for each instrument, the Data Processing Board will receive and distribute PPS signals
from the spacecraft to the instrument BEEs. 
 The Spacewire protocol is used for communications between the DPU and the S/C. 
 One independent board will be dedicated to the power supply: it will receive the primary power from the spacecraft and latch it through secondary power lines to all  instruments’ BEE units and to the alert transponder. Protection will be implemented against predefined voltage and current limits.  Moreover two additional boards will handle the TM/TC communication with the HXI and TED instruments BEEs. Besides implementing functions for the acquisition of the scientific data they will provide the necessary control sequences for configuration and operation of the instruments.
\begin{figure}[t]
  \begin{center}
  \includegraphics[height=7cm]{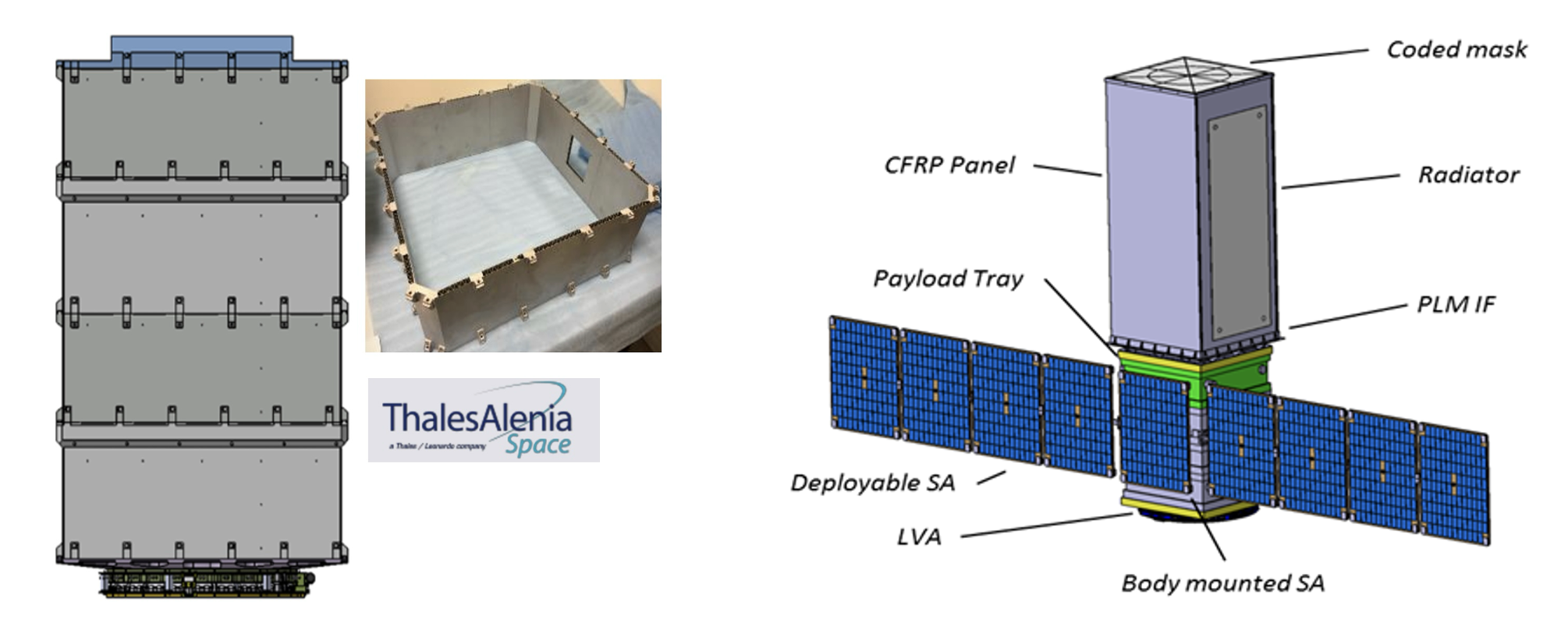}
   \caption{\textit{GRINTA} Spacecraft Overview. (Left) schematic of the NIMBUS trays structure (adapted from Ref.\citenum{TASI2024}). (Right) overview of the GRINTA satellite.}
    \label{fig:grinta_sc_overview}
      \end{center}
\end{figure}

No active cooling is foreseen with dissipation of the heat load to be obtained by conduction through mechanical interfaces with the satellite structure.

\section{Spacecraft design}
Driven by requirements for rapid repointing and a mass-constrained design, the GRINTA mission adopts the NIMBUS modular small satellite platform as its baseline architecture. Developed by Thales Alenia Space-Italia (Rome), NIMBUS\cite{TASI2024} satisfies mission requirements regarding performance, hardware maturity, and cost efficiency. The platform utilises a modular layout composed of standard, lightweight subsystems
“trays” (e.g., EPS, Data Handling, AOCS) and optional customised modules (Fig. 9, left). These trays are produced via additive manufacturing, serving simultaneously as structural elements and electronic enclosures to minimise dry mass. The core platform is qualified for "Extended LEO", 300–1200 km altitude with 0~deg inclination to SSO and achieves fast repointing via a miniaturised CMG housed in the AOCS tray.

\begin{figure}t]
  \centering
   \includegraphics[height=12cm, angle=-90]{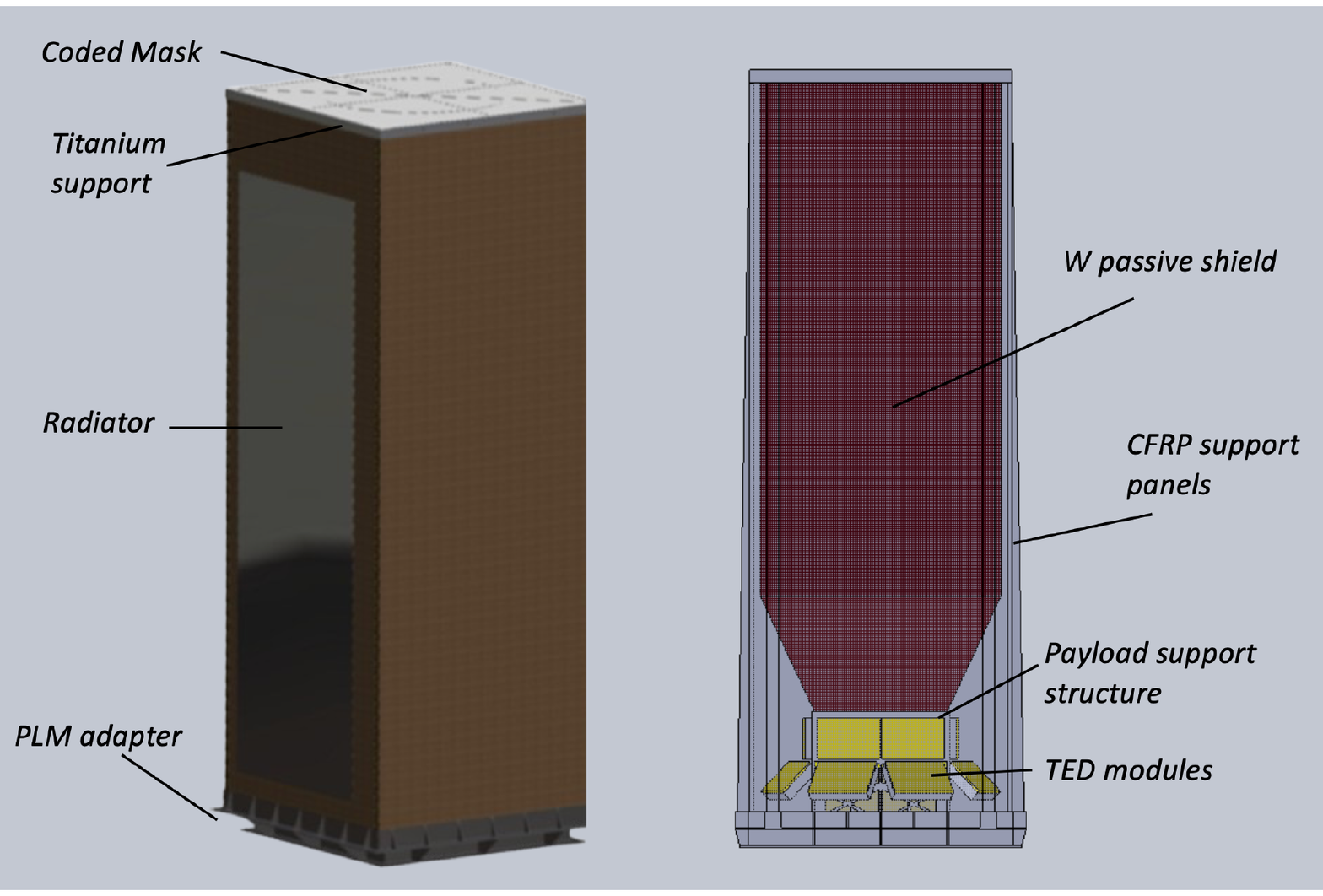}
   \caption{\textit{GRINTA} PLM Overview.}
    \label{fig:grinta_plm}
\end{figure}

The spacecraft bus is partitioned into a Payload Module (PLM) and a Service Module (SVM). The PLM consists of a rectangular structural enclosure that supports the scientific instruments and the coded mask 
(Fig.~\ref{fig:grinta_plm}). At its base, a monolithic, CNC-milled Aluminum 7075 plate provides the mechanical interface between the instrument suite and the NIMBUS SVM. This baseplate mounts the detector support structure, which centers the HXI detector (HXD) plane and positions the TED modules. Atop this frame, a hopper-shaped tungsten shield minimises stray radiation on the HXD without obstructing the TED field-of-view (FoV).

Integral mounting flanges on the baseplate facilitate bolted connections to the lateral panels. These 1800~mm long, 15~mm thick panels feature a honeycomb core with Carbon Fiber Reinforced Polymer (CFRP) skins, assembled via cleats and inserts to provide the structural rigidity required to support the upper coded mask. A 40~$\mu\text{m}$ tungsten lining is bonded to the internal faces of the panels to shield the instrumentation from low-energy photons and lower the detector background. Radiator plates for instrument thermal control are mounted externally on two lateral panels. The coded mask attaches to the PLM via a dedicated titanium support consisting of a square interface bracket secured to the honeycomb panels and a circularly reinforced, thin-ribbed central cross-structure designed to minimise HXI FoV obscuration.

The SVM employs a Multi-Functional Modular Frame (2-MF) approach, stacking a variable number of standardised trays to adapt to specific internal and external interface requirements. This structural subsystem provides the mechanical margins required for compatibility with most commercial launch vehicles, alongside integrated radiation shielding, an electrical Faraday cage, and an equipotential external surface. The baseline GRINTA configuration comprises four stacked trays, schematised in Fig.~\ref{fig:grinta_sc_overview}:


\begin{itemize}[noitemsep, leftmargin=*]
    \item bottom tray – Launch Vehicle Adapter (LVA) I/F, On-Board Data Handling (OBDH), and electric propulsion;
    \item intermediate tray 1 – AOCS reaction wheels;
    \item intermediate tray 2 - EPS subsystem and AOCS magnetorquers; 
    \item top tray: instrument support and BEEs
\end{itemize}


\section{Mission scenario}

\subsection{Reference Orbit}
The \textit{GRINTA} satellite will operate from a relative stable circular LEO equatorial orbit of 600~km altitude, with an inclination of $\sim$\(5^{\circ}\) over the Earth’s equator. Those characteristics represent a preliminary best compromise between the \textit{GRINTA} science mission objectives and the reference VEGA-C launch performances, especially in the case of a possible shared launch scenario. From such reference orbit, the \textit{GRINTA} satellite have an orbital period of 1.6 hrs, without significant variation in its orbital parameters.

From a preliminary study of the \textit{GRINTA} orbit stability, taking into account the Earth’s J2 effect and the drag aerodynamic force, as well as the solar radiation pressure, it is expected that a natural decaying of the \textit{GRINTA} orbit altitude, leading to an autonomous atmospheric re-entry of the satellite will occur in $\sim10$ years from the launch.

\subsection{Communication Architecture}
From \textit{GRINTA} science operation standpoint, relying only on terrestrial line-of-sight ground stations will result in a limitation from an operational and performance point of view. Considering the \textit{GRINTA} satellite autonomous re-pointing capability, 
communication crosslink using existing commercial networks is one of the most attractive and valuable option. It represents a clear advantage to remove the need of developing and maintaining new ground stations, 
resulting in positive effects on the operational costs of the mission. 

An average value of daily access time of about 8\% of the \textit{GRINTA} orbital period could be covered by operation of two ground stations. This figure corresponds to a maximum of $\sim2.1~hrs$ of total daily coverage.
Iridium~NEXT allows bi-directional link and convenient H/W  solutions to implement effective communication with its network. For its high-inclined satellites constellation, a maximum coverage time per day of –almost- 16~hours is estimated, corresponding an average value of daily access time about 66\% of \textit{GRINTA} orbital period, with a related minimum average value of around 27\%, corresponding to a daily overall minimum access time of 6.5~hours. 

 The ``Short Burst Data" (SBD) service provided by Iridium~NEXT allows efficient transmission of short messages and can be handled by purchasing a service plan at low cost, see e.g. https://www.groundcontrol.com/. Since a decade ago, Iridium communication devices have been proven in flight on board cubesats  within LEO. Different experiments have shown that these commercial transceivers (like the models Iridium SBD~9602 and the more recent SBD~9603) can receive commands and send low-volume telemetry using the Iridium SBD protocol. This communication system also demonstrated capability of handling the high relative velocities ($\sim7-10$~km/s) and resulting Doppler shifts in LEO. In December 2019, the NASA/LLNL MiniCarb satellite \cite{Wilson2021} successfully cross linked with Iridium~NEXT from LEO using a SBD~9523 transceiver during 72 hours (prior to S/C failure). They reported that for 90\% of communication attempts the delay between message generation and TLM reception was less than 30min.
A transceiver, e.g. a SBD~9602 or 9603 module, will be  mounted on the satellite and controlled by the \textit{GRINTA} DPU through power and serial data interface(s). A RHCP antenna designed for L-Band can be used, and it can be either a small helix antenna or a patch antenna.

\subsection{Ground Segment Elements}
TM/TC services will be provided by two ESTRACK ground stations (Kourou and, optionally, Malindi), managed by the ESA Mission Operations Centre (MOC). The MOC will oversee daily health monitoring, maintenance, and
anomaly resolution, alongside core flight dynamics functions, including launch window calculation, orbit and trajectory determination, debris tracking, and orbital maneuver execution. The ESA-hosted Science
Operations Centre (SOC) will manage instrument operations, scientific mission planning, quick-look analysis, and public data archiving. The SOC will interface with the scientific community to support the General Observer (GO) program by receiving and technically evaluating observation proposals. Optimised observation plans will then be forwarded to the MOC for conversion into spacecraft uplink telecommands.

The Consortium-led \textit{GRINTA} Science Data Centre (GSDC) will receive science and HK data from the SOC. The GSDC is responsible for hardware infrastructure, data processing, data product generation, analysis tools, storage, and comprehensive documentation. Both centres will utilise instrument-specific software and data, such as calibration files (e.g., energy response matrices), instrument configurations and hardware settings, provided by the instrument teams. Optionally, the Consortium may provide operational pipelines to support the SOC.

Closely integrated with the science ground segment, a Consortium-managed Burst Alert Center (BAC) will process rapid alerts from \textit{GRINTA}/TED. The BAC will distribute trigger data to the SOC and MOC, and disseminate alert notices across standard networks (e.g., GCN). It will also forward external alerts from ground-based observatories, such as GW and neutrino networks, and space-based assets. To enable autonomous, rapid satellite repointing, an automated system will be implemented at the SOC and MOC to handle GW alerts. For non-GW alerts, designated burst advocates will manually screen for false positives.

\section{\textit{GRINTA} Observing Strategy}
The main science goals of \textit{GRINTA} are focused on observing GRBs and EM counterparts to MM events, which come from random locations on the sky.  Thus the observing strategy while in survey mode is to observe semi-random positions for a roughly even exposure time of the full sky every few days, as is done with \textit{Swift} \cite{Tueller2010}.  During survey mode, it will be required that for most of the time the angle between the S/C boresight and the Earth zenith is kept lower that a given value (e.g. $\lesssim45^{\circ}$)  to optimise the sky viewing from the TED detectors. This strategy will produce a hard X-ray survey of the sky while searching for bursts.  

When TED detects and localises an event, \textit{GRINTA} will repoint autonomously to have the location within the HXI FoV for follow-up observations to search for afterglow emission and measure a more precise location of the event. These observations are expected to last $\sim10^4$~s.

Another possible source of triggers are alerts for transients received from ground. Potentially, these could be generated by ground facilities like the GW detectors, or by space observatories with capability to send immediately alerts with sky positions.  In this scenario, \textit{GRINTA} could autonomously repoint to the reported locations upon receiving the alert, provided certain conditions are met. 

\begin{table}[h!]
    \centering
    \caption{\textit{GRINTA} Yearly Time Allocation (example).}
    \label{tab:annual_program}
    \small
    \begin{tabular}{|p{2.94cm} |p{2.4cm} |p{1.2cm} |p{3.14cm} |p{2.0cm} |p{3.0cm} |}
    \hline
    \textbf{Scientific Topic} & \textbf{Obs. Type} & \textbf{Ms/yr} & \textbf{Obs. Mode} & \textbf{Alerts} & \textbf{Synergies} \\
    \hline
    \multicolumn{6}{c}{\textbf{--- Core Multi-Messenger Program ---}}\\
    \hline
        Gamma-Ray Bursts & ToO / Follow-up & 4.0 & Autonomous repointing & To Ground & ELT, CTAO, \textit{newATHENA}, etc \\
        Gravitational Waves & ToO / Follow-up & 5.0 & Repointing (on Ground alert) & from Ground & LVK, ET, CE \\
        Ext. Triggered Transients & ToO / Follow-up & 0.5 & Repointing (on Ground alert) & from Ground & Multi-wavelength \\
        \hline
                \multicolumn{6}{c}{\textbf{--- Survey \& General Science ---}}\\
                \hline
        Galactic \& Extragalactic surveys & Survey & 11.0 & Semi-random pointings /pointed obs. & to Ground & \textit{newATHENA}, SKAO, Vera Rubin  \\
        FRB \& Fast Transients & Survey/ToO & 3.0 & Semi-random pointings & to Ground & SKAO, CHORD, DSA-2000, Vera Rubin, ELT \\
        UHECR \& Neutrinos & Survey/ToO & 2.0 & Semi-random pointings / pointed obs. & Both & IceCube, KM3NeT \\
        \hline
 \textbf{Total Time} & & \textbf{25.5} & & & \\
        \hline
\end{tabular}
\end{table}

Based on the number of expected GRB detections per year from TED ($\approx385$), there will be approximately one GRB repointing per day, and routinely $\sim3-4$ repointing maneuvers per orbit to observe pre-determined fields in survey mode.
Assuming a duty cycle of 85\% and $\sim0.5$Ms/yr to be reserved for in-flight calibrations, the total amount of observing time available each year is 26\,Ms. Approximately 4\,Ms of observing time will be devoted to follow-up observations (10\(^4\)\,s/event).  $\sim4$~GW/EM joint detections are predicted for the 3$A^{\#}$ scenario and it can be anticipated that \(\sim\)5\,Ms in total exposure time will go towards follow-up observations for these events.  
 
Besides follow-up observation, $\sim$half of the available \textit{GRINTA} observing time (see Table~\ref{tab:annual_program}) will be devoted to surveys. Semi-random pointings will be programmed to cover the widest possible area of the sky while preserving efficient sky viewing of the TED modules. This will allow an efficient survey of the extragalactic populations. Other observations could be devoted to observe pre-determined sky regions to study and monitor galactic sources, candidate neutrino \& FRB sources, etc. 

The above observing strategy should be regarded as a preliminary assessment. The optimisation of the \textit{GRINTA} operability concept will be further refined during the forthcoming study phase and later, taking into account the scientific priorities that prevail at the time of the mission’s launch.


\section{Conclusions}
\textit{GRINTA} will provide a breakthrough impact on multimessenger and time domain astrophysics of the next decade by observing explosive events from the most remote regions of the Universe. Through rapid repointing with its HXI instrument, accurate event localisation and immediate transmission of alerts  will allow efficient follow-up with IR/optical/UV telescopes. Measurements of redshifts by IR/optical follow-up of sGRBs seen in coincidence with gravitational wave will have impact on cosmology (H0 measurements) and fundamental physics (e.g. theories of modified gravity) and will also allow to investigate the relative contribution of mergers and core-collapse SNae to the r-process (connection to cosmic-ray science, origin of heavy elements).
\textit{GRINTA} will also perform investigation of alerts generated on ground by radio, optical and VHE (e.g. SKAO, Vera Rubin, ELT,  CTAO), including subthreshold searches (events like e.g. GRBs, TDEs, FRBs) and search
for HE neutrino counterparts in the error regions of neutrino telescopes: IceCube Gen2, KM3NeT. Moreover, its
sensitive surveys will allow to detect (and follow-up for a subset of them) many astrophysical transients and
characterise thousands of hard X-ray sources. 

\acknowledgments 
J.Ripa, FM, NW thank the support by the Czech Science Foundation (GAČR) project No. 24-11487J. LH, AU and DM acknowledge support from Research Ireland grant 19/FFP/6777. JMMH and JAG are funded by Spanish MICIU/AEI/10.13039/501100011033 and ERDF/EU grant PID2023-147338NB-C21. PB acknowledges grant supported by Italian Research Center on High Performance Computing Big Data and Quantum Computing (ICSC), CN 00000013-CUP C53C22000350006. LN, GB acknowkledge funding from the EU Horizon 2020 Programme (AHEAD2020, grant agreement n. 871158).



\bibliographystyle{spiebib} 
\bibliography{sample631}
\end{document}